\documentclass[%
twocolumn,
superscriptaddress,
 amsmath,amssymb,
 aps,
prl,
]{revtex4-2}

\usepackage{xcolor} 
\usepackage{graphicx}
\usepackage{dcolumn}
\usepackage{bm}

\usepackage{color}

\usepackage{titlesec}
\usepackage{textcase}  
\usepackage{mfirstuc}  

\usepackage{hyperref}

\makeatletter
\providecommand{\subsubsection}{}
\renewcommand{\subsubsection}{%
  \@startsection{subsubsection}{3}{0pt}%
    {1.0ex plus .2ex minus .2ex}
    {0.5ex plus .2ex}
    {\normalfont\normalsize\bfseries\filcenter}%
}
\makeatother

\renewcommand{\thesection}{\arabic{section}}
\renewcommand{\thesubsection}{\thesection.\arabic{subsection}}
\renewcommand{\thesubsubsection}{\thesubsection.\arabic{subsubsection}}

\titleformat{\section}
  {\normalfont\large\bfseries\filcenter}
  {\thesection}{1em}{\MakeTextUppercase}

\titleformat{\subsection}
  {\normalfont\normalsize\bfseries\filcenter}
  {\thesubsection}{1em}{\capitalisewords}

\titleformat{\subsubsection}
  {\normalfont\normalsize\bfseries\filcenter}
  {\thesubsubsection}{1em}{\capitalizethefirst}

\usepackage{amsmath,amssymb,bm,amsthm}
\usepackage{mathtools}
\usepackage{multirow}
\usepackage{textcomp}
\usepackage{float}
\usepackage{xcolor}
\usepackage[normalem]{ulem}
\usepackage{dsfont}
\usepackage{mathrsfs}

\usepackage{tikz}
\usetikzlibrary{shapes, arrows, positioning}

\newcommand{\ket}[1]{\left | #1 \right \rangle}

\newcommand{\opa}{\hat{a}}
\newcommand{\opad}{\hat{a}^\dagger}
\newcommand{\opb}{\hat{b}}
\newcommand{\opbd}{\hat{b}^\dagger}
\newcommand{\opc}{\hat{c}}
\newcommand{\opcd}{\hat{c}^\dagger}

\newcommand{\oph}{\hat{H}}

\newcommand{\ddt}{\frac{\text{d}}{\text{d}t}}

\begin{document}

\preprint{APS/123-QED}


\title{Experimental Realization of Synthetic Magnonic Lattice via Floquet Engineering}

\author{Amin Pishehvar}
\affiliation{ 
    Department of Electrical and Computer Engineering, Northeastern University, Boston, MA 02115, USA
}

\author{Jayakrishnan M. P. Nair}
\affiliation{ 
    Department of Physics, Boston College, 140 Commonwealth Avenue, Chestnut Hill, MA, 02467, USA
}

\author{Zhaoyou Wang}
\affiliation{ 
    Pritzker School of Molecular Engineering, University of Chicago, Chicago, IL 60637, USA
}

\author{Zixin Yan}
\affiliation{ 
    Department of Electrical and Computer Engineering, Northeastern University, Boston, MA 02115, USA
}

\author{Yu Jiang}
\affiliation{ 
    Department of Electrical and Computer Engineering, Northeastern University, Boston, MA 02115, USA
}

\author{Liang Jiang}
\affiliation{ 
    Pritzker School of Molecular Engineering, University of Chicago, Chicago, IL 60637, USA
}

\author{Benedetta Flebus}
\affiliation{ 
    Department of Physics, Boston College, 140 Commonwealth Avenue, Chestnut Hill, MA, 02467, USA
}

\author{Xufeng Zhang}
\email{xu.zhang@northeastern.edu}
\affiliation{ 
    Department of Electrical and Computer Engineering, Northeastern University, Boston, MA 02115, USA
}
\affiliation{ 
    Department of Physics, Northeastern University, Boston, MA 02115, USA
}

\date{\today}

\begin{abstract}
Magnonic systems, which exploit spin-wave excitations in magnetic materials, offer a promising platform for coherent information processing due to their low dissipation, strong nonlinearities, and intrinsic nonreciprocity. However, scaling magnonic circuits remains challenging, particularly with low-loss insulators such as yttrium iron garnet (YIG), which are difficult to pattern. Here, we experimentally realize a synthetic dimension in a magnonic system by coupling multimode magnon resonances in the frequency domain using time-periodic Floquet modulation. This approach enables electronically tunable interactions between discrete modes within a single YIG device, forming a reconfigurable mode-space lattice that supports functionalities such as Bloch oscillation. Our results demonstrate that high-dimensional magnonic dynamics can be achieved without increasing device footprint, establishing synthetic dimensions as a scalable and programmable route for integrated magnonic technologies. This advancement positions magnonic systems as promising platforms for engineering emergent phenomena that are inaccessible at equilibrium.
\end{abstract}


\maketitle



Magnonics--the study and manipulation of spin waves in magnetic materials--has emerged as a promising platform for coherent information processing \cite{HybridmagnonicsLi,QuantumEngineeringWithHybridMagnonic,BHOI202039,HARDER201847,Lachance-Quirion_2019, Chumak2019MagnonSpintronics, flebus_2024_2024,ZARERAMESHTI20221,Areviewofcommonmaterials,StronglyCoupledMagnons,StrongFieldInteractionsSoykal,High-CooperativityCavity,Magneto-opticalcouplingHaigh,Cavitymagnomechanics}. As collective excitations of magnetization, magnons exhibit unique properties such as immunity to ohmic losses, strong nonlinear interactions, and intrinsic nonreciprocity. These features make them attractive for applications in computing \cite{wang_magnonic_2020,zenbaa_universal_2025,AdvancesinMagneticsRoadmap,meta-learningZhang,korber_pattern_2023,yaremkevich_-chip_2023}, sensing \cite{flower_broadening_2019,AxionSearchCrescini,DetectingLightDarkMatterwithMagnons,chang_mosaic_2025,barry_ferrimagnetic_2023,matatagui_magnonic_2017,silva_biosensor_2024}, and quantum technologies \cite{HighCooperativityHuebl,Resolvingquantaofcollectivespin,Entanglement-basedsingle-shot,HybridizingFerromagneticMagnonsTabuchi,CoherentCouplingTabuchi,QuantumControlXu,Dissipation-BasedWolski}. Recent advances in device fabrication \cite{rashedi_photonic_2025,MonocrystallineFreestandingHeyroth,trempler_integration_2020,baity_strong_2021} and the demonstration of integrated magnonic components, including logic gates \cite{inverse-designZenbaa}, transistors \cite{chumak_magnon_2014}, and reconfigurable circuits \cite{zenbaa_universal_2025}, have highlighted its potential as a complementary technology to electronics and photonics, particularly for compact and energy-efficient signal processing with unprecedented in-situ tunability.

Despite this progress, scaling magnonic systems remains a major challenge. The construction of large-scale, high-quality magnonic circuits is limited by material and fabrication constraints. Magnetic alloys are easy to pattern but suffer from high damping, while low-loss magnetic insulators such as yttrium iron garnet (YIG) have complex crystalline structures and are highly resilient to chemical etching, making them difficult to pattern using conventional techniques. These limitations hinder the development of scalable architectures and restrict access to advanced functionalities such as programmable interactions, topological transport, and many-body dynamics.

To address these challenges, we implement a novel synthetic dimension approach in magnonics. Synthetic dimensions \cite{yuan_synthetic_2018,SynthDimePerspectMaxErhardt} have recently been recognized as an effective framework for simulating higher-dimensional physics through the use of internal degrees of freedom such as frequency \cite{yuan_photonic_2016,yang_non-abelian_2025,SyntheticNon-AbelianWong}, polarization \cite{polarizationSynthDimeMaxEhrhardt}, and angular momentum \cite{luo_quantum_2015}. This concept has been successfully demonstrated in photonics \cite{yuan_synthetic_2018,SynthDimePerspectMaxErhardt,yang_non-abelian_2025,SyntheticNon-AbelianWong,polarizationSynthDimeMaxEhrhardt,luo_quantum_2015,Dutt_Science_2020}, ultracold atoms \cite{mancini_observation_2015}, and superconducting circuits \cite{Lee_PRA_2020,xiang_simulating_2023}, enabling compact realizations of high-dimensional topological phases and programmable interactions. Recently, the concept of synthetic frequency dimension has also been theoretically proposed for magnonic systems \cite{Xu2025_NC_magnonSynthDim}. In this work, we construct a synthetic dimension using high-order magnon modes in the frequency domain, combined with time-periodic modulation techniques developed in Floquet magnonics \cite{FloquetCavityElectromagnonics,On-demandmagnonresonanceisolation,Pishehvar_PRAppl_2025_resonance}. This enables electronically tunable interactions between discrete magnon modes, forming a reconfigurable mode-space lattice within a single YIG device that supports functionalities such as Bloch oscillation. To the best of our knowledge, this is the first experimental realization of a synthetic dimension in a magnonic system. Our results demonstrate that complex, high-dimensional magnonic dynamics can be achieved within a compact device, establishing synthetic dimensions as a scalable and programmable route for integrated magnonic technologies.

\begin{figure}[tb]
\includegraphics[width=0.99\linewidth]{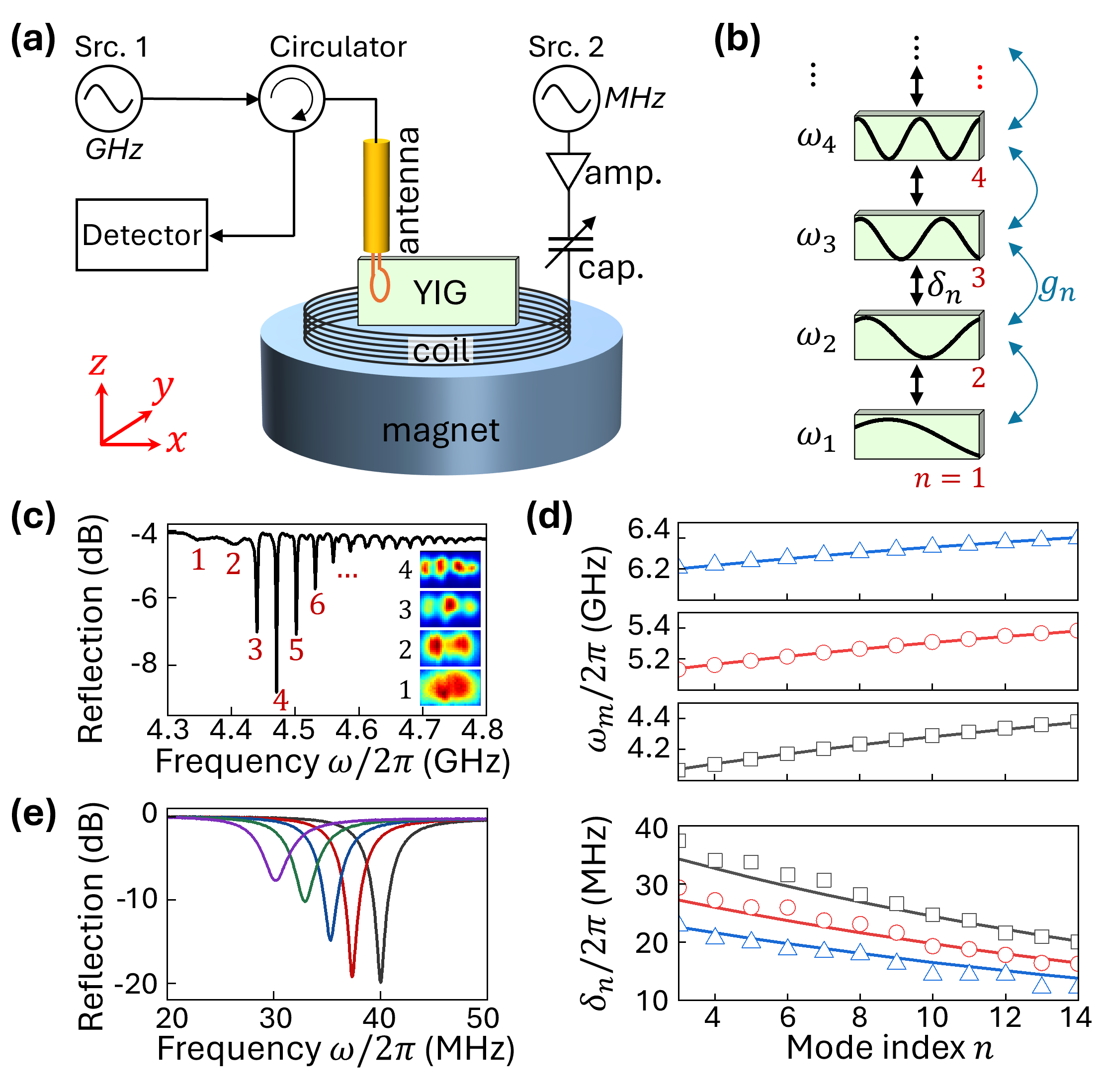}
\caption{(a) Schematic of the magnonic synthetic dimension device and measurement setup. (b) Conceptual illustration of the spectral synthetic dimension: discrete magnon resonances separated by $\delta_n$ and coupled via Floquet-induced strengths $g_n$. (c) Measured reflection spectrum showing magnon modes indexed by $n = 1, 2, 3, \dots$. Inset: Measured mode profiles. (d) Extracted magnon frequency $\omega_m$ and mode spacing $\delta_n$ versus mode index $n$ under three bias conditions. (e) Measured LC resonances as the tunable capacitor is varied. Src.: source; cap.: capacitor; amp.: amplifier.}
\label{fig1}
\end{figure}


The schematic of our experimental setup is shown in Fig.\,\ref{fig1}(a). The magnonic device is a $23~\mu \mathrm{m} \times 2~ \mathrm{mm} \times 3~\mathrm{mm}$ single-crystalline YIG film, biased by an external magnetic field parallel to its short edge (along $z$). Due to the finite chip size, magnons form a series of discrete standing wave modes [Fig.\,\ref{fig1}(b)] along $x$ direction, which can be accessed using a coaxial loop antenna. From the device reflection spectrum [Fig.\,\ref{fig1}(c)] measured using a vector network analyzer (VNA) at around 4.4 GHz, over 15 magnon resonances can be clearly observed. The spectrum shows that, as the frequency increases, these modes exhibit reduced mode spacings [Fig.\,\ref{fig1}(d)], suggesting that they are magnetostatic surface spin waves (MSSWs) in nature. This is further confirmed by the measured mode profiles in the inset of Fig.\,\ref{fig1}(c), which show up as standing wave modes along the length direction of the YIG chip, indicating that the dominant wavevector $k$ is perpendicular to the bias magnetic field direction. These magnon modes form a lattice in the frequency domain, which serves as the basis for the magnonic synthetic dimension.

To utilize the synthetic dimension for coherent signal processing, it is essential to achieve controllable coupling between these discrete magnon modes. Floquet engineering has recently been demonstrated as an effective method for inducing such coupling in hybrid magnonic systems, enabling phenomena such as Autler–Townes splitting  \cite{FloquetCavityElectromagnonics}, on-demand mode isolation \cite{On-demandmagnonresonanceisolation}, and selective mode coupling \cite{Li2023FloquetMagnon}. Here, we apply a single Floquet drive to the magnon mode lattice, extending the Floquet magnonics to a previously unachievable high-dimensional setting, but yet on a single device and thus without significantly increasing the device size and complexity. The Floquet drive from a MHz source is amplified and then sent to a compact coil, which is aligned with the external bias field. It generates a time-periodic modulation of the magnetic field and, consequently, the magnon resonance frequencies. When the drive frequency is close to the spacing between adjacent magnon modes in the lattice, inter-mode coupling is enabled, allowing coherent energy transfer between modes. To enhance the modulation strength, the coil is connected in series with a variable capacitor, forming an LC resonator with a resonance frequency tunable between 20 and 40 MHz [Fig.\,\ref{fig1}(e)] to match the magnon mode spacing.


On our device, the dynamics of the magnetization \(\mathbf m(\mathbf r,t)\) under a static in-plane bias \(\mathbf H_0=H_0\hat{\mathbf z}\) and a weak longitudinal modulation \(H_1\cos(\Omega t)\,\hat{\mathbf z}\) can be described by the  Landau-Lifshitz-Gilbert (LLG) equation
\begin{equation}
\dot{\mathbf m}=-\gamma\,\mathbf m\times\!\big[\mathbf H_0+\mathbf h_d(\mathbf m)+H_1\cos(\Omega t)\,\hat{\mathbf z}\big]+\alpha\,\mathbf m\times\dot{\mathbf m},\label{llg}
\end{equation}
where $\gamma$ is the gyromagnetic ratio, $\alpha$ denotes the Gilbert damping, and $\mathbf{h}_d(\mathbf{m})$ represents the magnetic field generated by the dipolar interaction. Linearizing about the saturation magnetization $m_s\hat{\mathbf z}$ yields, within the framework of our LLG model, the multimode Floquet magnonic system can be described by a tight-binding Hamiltonian, which takes the following form in the frame rotating at $\omega_\mathrm{ref}+n\Omega$ (see Supplemental Material \cite{SM} for details):
\begin{equation}
    \hat{H}=\sum_{n=1}^{N}\Delta_n\hat{a}_n^\dagger\hat{a}_n+\sum_{n=1}^{N-1}(g_n\hat{a}_n^\dagger \hat{a}_{n+1}+h.c.),
\label{Eq:Hamiltonian_RWA}
\end{equation}
\noindent 
where $\hat{a}_n^\dagger$ ($\hat{a}_n$) is the creation (annihilation) operator of the $n$-th magnon mode, $\Delta_n=\omega_n-n\Omega-\omega_\mathrm{ref}$ is the detuning of each mode from the Floquet lattice in the rotating frame with respect to the reference $\omega_\mathrm{ref}$, $\omega_n$ is the $n$-th magnon mode in the lab frame, $N$ is the total number of modes, $g_n$ is the coupling strength between the $n$-th and $(n+1)$-th mode, and $\Omega$ is the Floquet drive frequency, respectively. The reference frequency $\omega_\mathrm{ref}$ is typically selected at one of the magnon mode frequencies. Note that to capture the essential dynamics while maintaining analytical tractability, we consider only nearest-neighbor inter-mode coupling \cite{SM}. The interaction strengths $g_n$ can be obtained by solving Eq.\,(\ref{llg}) using the Floquet analysis and are approximately given by
\begin{equation}
    g_n\approx \frac{\gamma m_s}{2}\tilde{G}_{n,n+1}\mathcal{J}_1(\Delta/\Omega),
\end{equation}
where $\Delta$ is the magnitude of the magnon frequency deviation induced by the Floquet drive, $\mathcal{J}_1$ is the Bessel function of order one, and $\tilde{G}_{n,n+1}$ is the overlap integral between the spatial mode profiles of modes $n$ and $n+1$ (see Supplemental Material  \cite{SM} for details). 


One distinctive advantage of our magnonic synthetic dimension is its high tunability. Magnon modes exhibit strong frequency dependence on the bias magnetic field, allowing the entire set of discrete modes to be continuously shifted across a wide frequency range [Fig.\,\ref{fig1}(d)]. In addition, the mode spacing also varies with the bias magnetic field, enabling on-demand tuning of the required Floquet drive frequency. As shown in the top panel of Fig.\,\ref{fig1}(d), the slope of the magnon dispersion decreases with frequency, leading to reduced mode spacing $\delta_n=\omega_{n+1}-\omega_n$ for higher-order modes, as plotted in the bottom panel of Fig.\,\ref{fig1}(d). For example, $\delta_{3}$ is nearly twice of $\delta_{13}$. Consequently, by properly choosing the Floquet drive frequency, it is possible to achieve selective mode coupling. Compared to previous demonstrations of synthetic dimensions in photonic, acoustic, and electronic platforms where mode lattices are fixed by device geometry and material properties, our system offers unprecedented reconfigurability and dynamic tunability.


\begin{figure}[tb]
\includegraphics[width=0.99\linewidth]{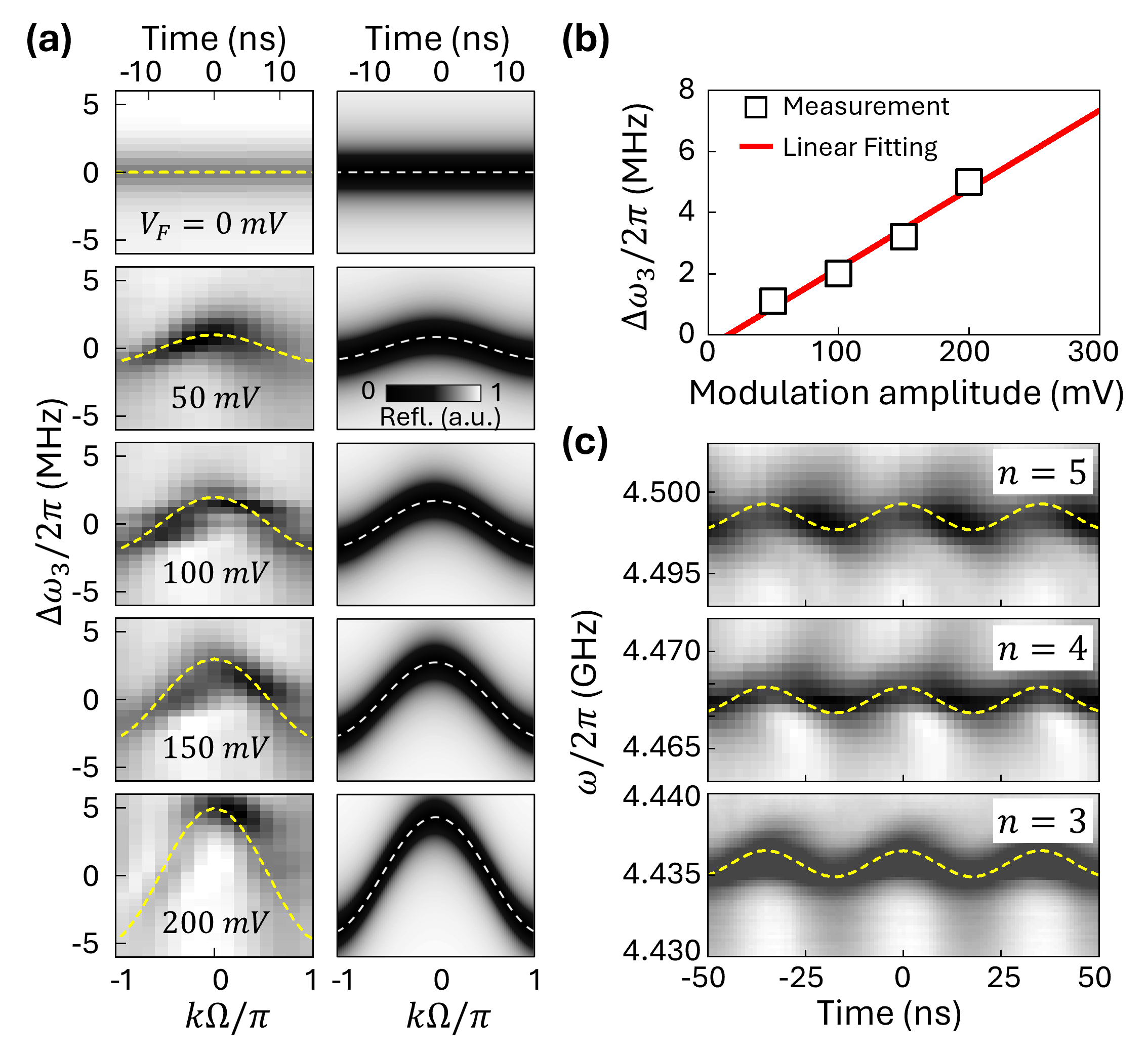}
\caption{(a) Measured (left) and calculated (right) reflection spectra of Mode 3 at various Floquet drive amplitudes $V_F$, shown as a function of time (top $x$ axis) and corresponding synthetic quasimomentum $k$ (bottom $x$ axis). $y$ axis: $\Delta\omega_3 = \omega'_3 - \omega_3$. (b) Extracted frequency deviation $\Delta\omega_3$ v.s. Floquet drive amplitude, representing modulation depth. (c) Time-domain oscillations of magnon modes with orders $n = 3, 4, 5$. Dashed lines in (a) and (c): cosine fits. Refl.: Reflection.}
\label{fig2}
\end{figure}

To characterize the properties of the magnonic synthetic dimension, we first performed time-resolved band structure spectroscopy \cite{dutt_experimental_2019} to map out the Floquet band structure, i.e., the dispersion relation between the quasienergy $\varepsilon$ and quasimomentum $k$, of our time-modulated magnonic system. Note that for the synthetic dimension in the frequency domain, the quasimomentum $k$ is conjugate to the frequency and lies along the time axis. Figure\,\ref{fig2}(a) shows the measured Floquet band for the third magnon mode (at $\omega_3$) under a modulation frequency $\Omega/2\pi = 28.4$ MHz, projected onto the first Brillouin zone $k \in [-\pi/\Omega, \pi/\Omega]$. Without modulation, the magnon mode frequency remains constant. As the drive amplitude increases from 50\,mV to 200\,mV (before amplification), the frequency deviation $\Delta\omega_3=\omega'_3 - \omega_3$, i.e., the difference between the modulated mode frequency $\omega'_3$ and the original frequency $\omega_3$, grows linearly, reflecting increasing quasienergy shifts [Fig.\,\ref{fig2}(b)]. This behavior is consistent with frequency modulation, where the magnetic field scales linearly with drive amplitude. Similar effects are observed across higher-order modes [Fig.\,\ref{fig2}(c)], confirming translational symmetry along the synthetic dimension. The yellow dashed lines in Figs.\,\ref{fig2}(a) and (c) represent theoretical predictions, showing excellent agreement with experiment.

\begin{figure}[tb]
\includegraphics[width=0.98\linewidth]{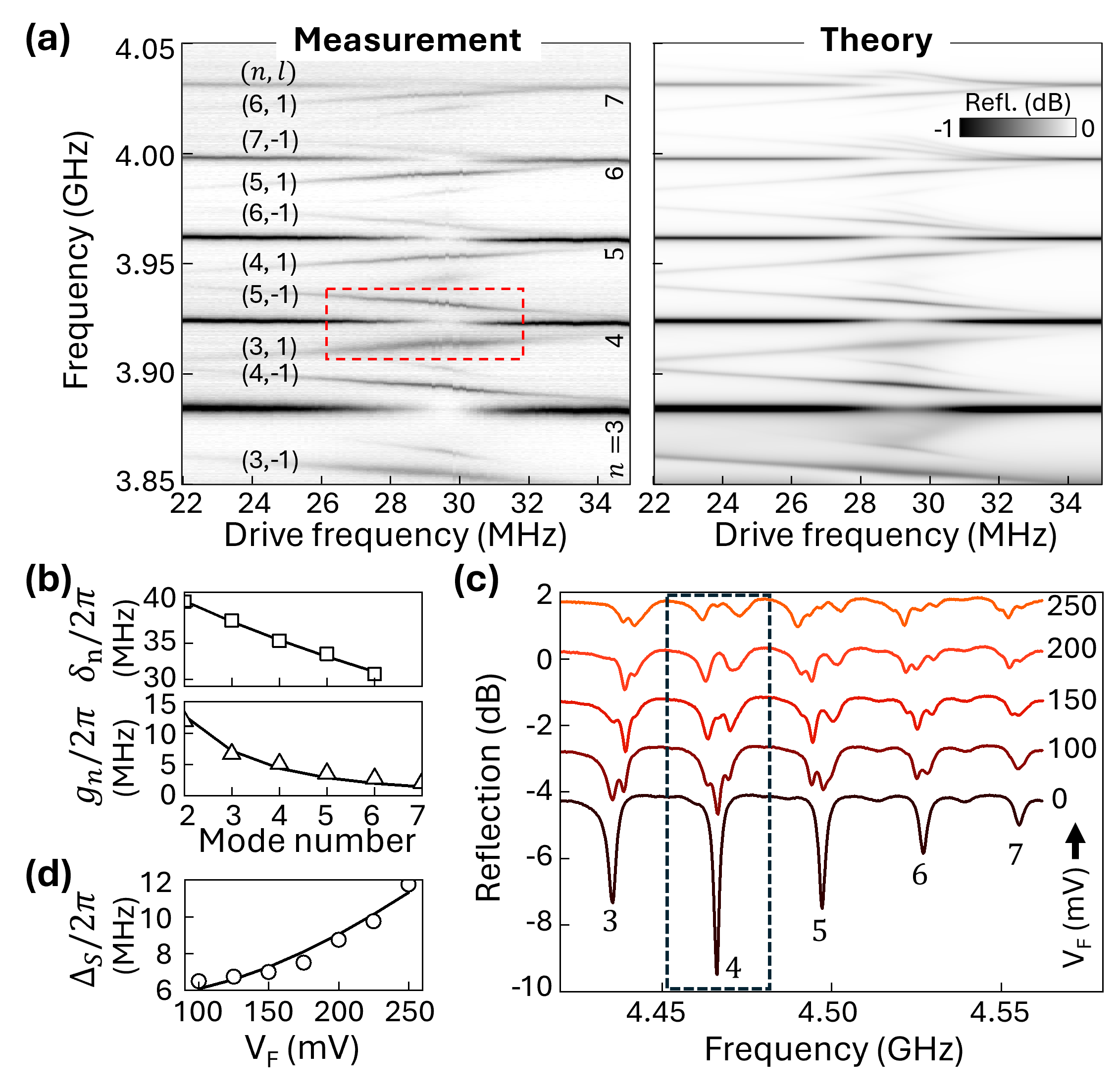}
\caption{(a) Measured and calculated steady-state reflection spectra, showing five modes in the synthetic dimension under a Floquet drive with $V_F = 300$ mV at different $\Omega/2\pi$. ($n$,$l$) represents the mode order ($n$) and sideband order $l$. (b) Mode spacing $\delta_n$ and coupling strength $g_n$ extracted from (a). Squares: experimental data; Triangles: extracted from experimental data through numerical calculation. Solid curves: micromagnetic calculation results. (c) Measured reflection spectra under varying drive amplitudes. (d) Floquet-induced mode splitting $\Delta_s$ for Mode 4 as a function of drive amplitude $V_F$, extracted from (c). Circles: experimental data: solid curve: micromagnetic calculation results.}
\label{fig3}
\end{figure}

To reveal the inter-mode coupling within the synthetic lattice, the steady-state reflection spectrum is measured [Fig.\,\ref{fig3}(a), left], which shows good agreement with numerical simulations based on Eq.\,(\ref{Eq:Hamiltonian_RWA}) [Fig.\,\ref{fig3}(a), right]. As the Floquet drive frequency is varied, each magnon mode exhibits multiple sidebands, labeled by their mode number $n$ and sideband order $l$. These sidebands become prominent when the drive frequency approaches the LC resonance at $2\pi\times 30$ MHz, which has a linewidth of approximately $2\pi\times 4$ MHz. Under this condition, the $n$-th magnon mode is simultaneously coupled to the ($n{-}1$, 1) and ($n{+}1$, $-1$) sidebands. These two sidebands form a bright mode which couples with the $n$-th mode, and a dark mode that remains decoupled from Mode~$n$ and is therefore not visible in the spectrum. This phenomenon is clearly illustrated by Mode 4 as an example, highlighted in the dashed box in Fig.\,\ref{fig3}(a), where the central mode gradually vanishes as the drive frequency scans across the LC resonance, resembling magnon dark modes observed under static coupling conditions \cite{Zhang2015MagnonDarkMode}.

The mode spacing $\delta_n$ extracted from Fig.\,\ref{fig3}(a) is plotted in the top panel of Fig.\,\ref{fig3}(b), showing a decreasing value with the mode index $n$, from 40 MHz for $n = 2$ to 32 MHz for $n = 6$. Although not all mode spacings can be simultaneously resonant with a given Floquet drive, a finite detuning does not eliminate Floquet-induced coupling, as shown in previous demonstrations \cite{FloquetCavityElectromagnonics, Pishehvar_PRAppl_2025_resonance}. With the detuning effect considered, the coupling strength $g_n$ is extracted from experimental data using our Hamiltonian model in Eq.\,(\ref{Eq:Hamiltonian_RWA}) and plotted in the bottom panel of Fig.\,\ref{fig3}(b). The coupling strength also shows a decreasing trend from 11 MHz to 2 MHz when $n$ increases from 2 to 7, which agrees well with our theoretical calculation (solid line) based on the spatial overlap between interacting modes.

The reflection spectra in Fig.\,\ref{fig3}(c) reveal how mode coupling depends on the Floquet drive amplitude. Without modulation ($V_F = 0$), magnon modes appear as non-perturbed Lorentzian dips. When a drive at $\Omega_F/2\pi = 27.5$\,MHz and $V_F = 100$\,mV is applied, the magnon extinction ratio decreases and two small kinks emerge on either side (see Mode 4 as an example), indicating coherent coupling to neighboring modes. As the drive amplitude exceeds 150\,mV, the single dip splits into two, signaling hybridization between the central mode (Mode 4) and a bright mode formed by the upper sideband of Mode 3 and the lower sideband of Mode 5. From the observed splitting $\Delta_s/2\pi = 7$\,MHz, the extracted coupling strength $g_{3}/2\pi$ is around 2.01\,MHz, which exceeds the dissipation rates of Mode 4 ($\kappa_4/2\pi=1.4$\,MHz) and the average dissipation of Mode 3 and Mode 5 [$\kappa_\mathrm{avg}/2\pi= 1.5$\,MHz], confirming the strong coupling condition. 
Increasing the drive amplitude further enhances the splitting [Fig.\,\ref{fig3}(d)], reaching up to $2\pi \times 12$\,MHz at 250\,mV, corresponding to $g_{3}/2\pi \approx 10.5$\,MHz and a cooperativity of $C = 53$. These results confirm coherent coupling between neighboring sites in the synthetic dimension, enabling robust information transfer across the magnonic mode lattice.


\begin{figure}[tb]
\includegraphics[width=0.99\linewidth]{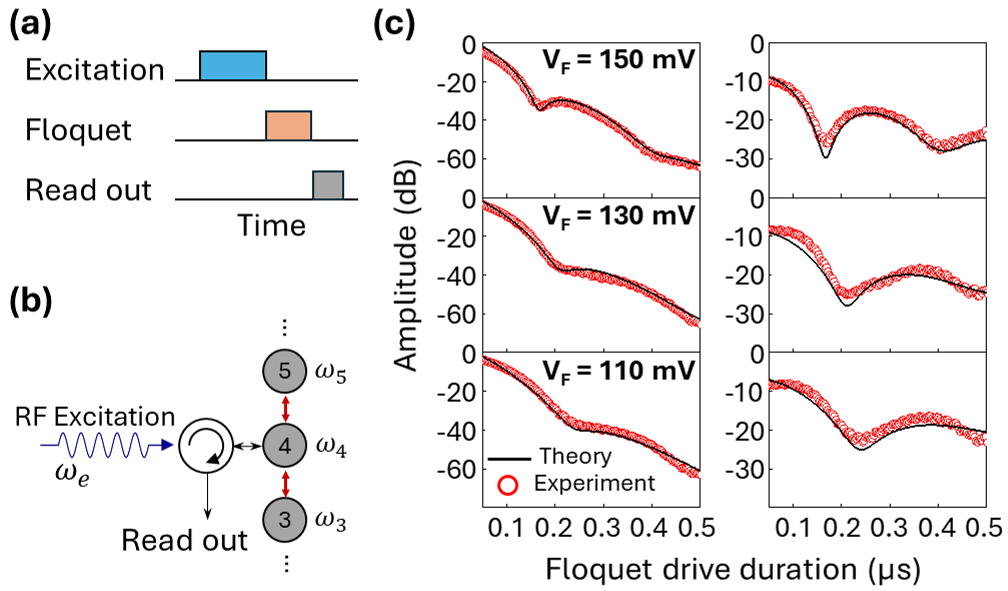}
\caption{(a) Schematic of the pulse sequence used for measuring spectral Bloch oscillations. (b) Illustration of the process: an RF pulse excites Mode 4 at $\omega_e/2\pi = \omega_4/2\pi=3.92$ GHz, and the Floquet drive induces coupling (red arrows) with neighboring modes, leading to energy exchange and oscillations in the measured power of Mode 4. (c) Left panel: time-resolved reflection signal from Mode 4 under a Floquet drive at $\Omega/2\pi=30$~MHz and varying amplitudes $V_F =$ 110, 130, and 150\,mV using the sequence in (a). Right panel: the same time traces after removing the exponential decay background. Solid lines are theoretical calculations using a single set of device parameters.}
\label{fig4}
\end{figure}

The Floquet-induced tunable mode coupling enables the observation of spectral Bloch oscillations, i.e., Bloch oscillation in the frequency domain, which is a hallmark phenomenon of synthetic dimensions that demonstrates coherent information exchange across discrete frequency modes. After Mode 4 is excited by a long RF pulse at the frequency $\omega_e/2\pi=\omega_4/2\pi=3.92$~GHz, a Floquet drive is applied, inducing the coupling of Mode 4 with the other modes in the lattice, as shown by the pulse sequence and schematics in Fig.\,\ref{fig4}(a)-(b). The resulting reflection spectra of Mode 4 for different modulation durations are plotted in Fig.\,\ref{fig4}(c). In the absence of Floquet drive ($V_F = 0$), a simple exponential decay is observed (see Supplemental Material \cite{SM}) due to magnon dissipation in Mode 4. Upon applying a Floquet drive, periodic oscillations can be unambiguously observed. The dependence of the oscillation period on drive strength is a finite-lattice effect. The results agree well with our numerical simulations, which are based on our model in Eq.\,(\ref{Eq:Hamiltonian_RWA}) using a single set of parameters for all four curves, with only the driving amplitude varied: $g_3/2\pi = 2.07$\,MHz, $g_4/2\pi = 1.15$\,MHz, $g_5/2\pi = 0.23$\,MHz, and $g_6/2\pi = 0.21$\,MHz, respectively. 

The nonuniform magnon mode spacing produces site-dependent potential gradients ($\delta_n - \Omega$), creating an effective quasi-linear potential across the locally coupled modes (Modes 3, 4, and 5), which supports the observed Bloch-type oscillatory dynamics (see Supplemental Material \cite{SM} for details). Clearly the coupling of Mode 4 with its nearest neighbors is comparable to the magnon dissipation rates ($\kappa_3/2\pi=1.70$ MHz, $\kappa_4/2\pi=1.29$ MHz, $\kappa_5/2\pi=1.35$ MHz), positioning the system in or near the strong coupling regime. Such Bloch oscillation indicates coherent information exchange between Mode 4 and adjacent modes in the magnon mode lattice, and equivalently, coherent signal propagation in the synthetic dimension. Note that the achievable Floquet-induced coupling can be further enhanced if the driving efficiency can be improved with more advanced device designs such as using superconducting circuits.


In summary, we have experimentally realized, for the first time, a magnonic synthetic dimension by coherently coupling discrete standing wave magnon modes in a YIG device via Floquet engineering. Time-periodic modulation of the bias magnetic field enabled tunable inter-mode coupling, forming a reconfigurable mode-space lattice with well-defined boundaries. Time-resolved spectroscopy revealed quasienergy dispersion and coherent dynamics across the synthetic dimension, including strong coupling and magnon dark states. Notably, we observed spectral Bloch oscillations, which is a hallmark of coherent control in synthetic dimensions, highlighting the potential for advanced magnon-based signal processing. To the best of our knowledge, this is the first experimental realization of spectral Bloch oscillation in magnonic systems. Compared to existing synthetic dimensions demonstrated in photonic, acoustic, or electronic platforms, our magnonic synthetic dimension system offers unique advantages such as large frequency tunability and adjustable mode spacing. These features position magnonic synthetic dimensions as a scalable and programmable framework for exploring emergent phenomena with no equilibrium counterpart and developing reconfigurable integrated magnonic circuits.

\section{Acknowledgments}
\begin{acknowledgments}

B. Flebus and X. Zhang acknowledge support from National Science Foundation under Grant No. NSF ECCS-2337713. Z.W and L.J acknowledge support from the AFOSR MURI (FA9550-21-1-0209, FA9550-23-1-0338), NSF (OSI-2426975) and the Packard Foundation (2020-71479).
\end{acknowledgments}


%

\onecolumngrid
\clearpage
\setcounter{section}{0}
\setcounter{figure}{0}
\setcounter{equation}{0}
\renewcommand{\thefigure}{S\arabic{figure}}
\renewcommand{\theequation}{S\arabic{equation}}
\renewcommand{\thesection}{S\arabic{section}}
\renewcommand{\thesubsection}{\thesection.\arabic{subsection}}
\renewcommand{\thesubsubsection}{\thesubsection.\arabic{subsubsection}}
\begin{center}
\textbf{\large Supplemental Material for\\[2pt]
``Experimental Realization of Synthetic Magnonic Lattice via Floquet Engineering''}
\end{center}
\vspace{1em}

\section{Theoretical Models}

\subsection{Hamiltonian model}

\subsubsection{Hamiltonian model and reflection spectrum under a Floquet drive}
In the lab frame, the system Hamiltonian is
\begin{equation}
	\oph_S(t) = \sum_{n=1}^N \omega_n \opad_n \opa_n + 2 \cos (\Omega t) \sum_{n=1}^{N-1} (g_n\opad_n \opa_{n+1} + h.c.) .
\end{equation}
We assume the system-bath coupling is described by
\begin{equation}
	\oph_{int} = i \sqrt{\frac{\gamma}{2\pi}} \int_{-\infty}^{\infty} d \omega (\opad(\omega) \opc + \opa(\omega) \opcd) ,
\end{equation}
where $\opc = \sum_n \sqrt{\frac{\kappa_{e,n}}{\gamma}} \opa_n$ is the system operator that couples to the external drive line with coupling rate $\gamma$.

The Langevin equations are
\begin{equation}
	\ddt \opa_n = -i [\opa_n, \oph_S(t)] - \frac{\kappa_{i,n}}{2} \opa_n - \frac{1}{2} \sum_m \sqrt{\kappa_{e,n} \kappa_{e,m}} \opa_m - \sqrt{\kappa_{e,n}} \opa_{in} (t) ,
\end{equation}
and the output field is
\begin{equation}
	\opa_{out} (t) = \opa_{in} (t) + \sum_n \sqrt{\kappa_{e,n}} \opa_n (t) .
\end{equation}
To derive the reflection spectrum, we can write the Langevin equation in the matrix form
\begin{equation}
	\ddt \begin{pmatrix} \opa_1 \\ \vdots \\ \opa_N \end{pmatrix} = -i M(t) \begin{pmatrix} \opa_1 \\ \vdots \\ \opa_N \end{pmatrix} - 
	\begin{pmatrix} \sqrt{\kappa_{e,1}} \\ \vdots \\ \sqrt{\kappa_{e,N}} \end{pmatrix} \opa_{in} (t) ,
\end{equation}
where $M(t) = H(t) - i \frac{K}{2}$ and
\begin{equation}
	H(t) = \begin{pmatrix} \omega_1 & & 0 \\ & \ddots & \\ 0 & & \omega_N \end{pmatrix} + 2 \cos (\Omega t)
	\begin{pmatrix}
		0 & g_1 & & & \\
		g_1 & 0 & g_2 & \\
		& g_2 & 0 & g_3 \\
		& & \ddots & \ddots & \ddots \\
		& & & g_{N-2} & 0 & g_{N-1} \\
		& & & & g_{N-1} & 0
	\end{pmatrix}
\end{equation}
and
\begin{equation}
	K = \begin{pmatrix} \kappa_{i,1} & & 0 \\ & \ddots & \\ 0 & & \kappa_{i,N} \end{pmatrix} +
	\begin{pmatrix}
		\sqrt{\kappa_{e,1} \kappa_{e,1}} & \cdots & \sqrt{\kappa_{e,1} \kappa_{e,N}} \\
		\vdots &  & \vdots \\
		\sqrt{\kappa_{e,N} \kappa_{e,1}} & \cdots & \sqrt{\kappa_{e,N} \kappa_{e,N}}
	\end{pmatrix} .
\end{equation}

From the Fourier expansion
\begin{equation}
	M(t) = M_0 + M_1 e^{i\Omega t} + M_{-1} e^{-i\Omega t} ,
\end{equation}
we can write the Langevin equations in the frequency domain as
\begin{equation}
	-i\omega \bm{a} (\omega) = -i \sum_{k=-1}^{1} M_k \bm{a} (\omega + k \Omega) - 
	\begin{pmatrix} \sqrt{\kappa_{e,1}} \\ \vdots \\ \sqrt{\kappa_{e,N}} \end{pmatrix} \opa_{in} (\omega) .
\end{equation}
where $\bm{a} = (\opa_1,...,\opa_n)$.

We can truncate the Langevin equations to the $k$th order, which gives
\begin{equation}
	G (\omega)
	\begin{pmatrix}
		\bm{a} (\omega - k\Omega) \\
		\vdots \\
		\bm{a} (\omega) \\
		\vdots \\
		\bm{a} (\omega + k\Omega) 
	\end{pmatrix}
	= -\begin{pmatrix}
		\bm{0} \\
		\vdots \\
		\sqrt{\mathbf{K}_e} \\
		\vdots \\
		\bm{0}
	\end{pmatrix}
	\opa_{in} (\omega) ,
\end{equation}
where
\begin{equation}
	G (\omega) = 
	\begin{pmatrix}
		D_{\omega-k\Omega} & i M_1 & & & \\
		i M_{-1} & D_{\omega-(k-1)\Omega} & i M_1 & & \\
		& \ddots & \ddots & \ddots & \\
		& & i M_{-1} & D_{\omega+(k-1)\Omega} & i M_1 \\
		& & & i M_{-1} & D_{\omega+k\Omega} 
	\end{pmatrix}
\end{equation}
with $D_\omega = -i (\omega I_n - M_0)$, and $\mathbf{K}_e = (\kappa_{e,1},...,\kappa_{e,N})^T$. The reflection coefficient is therefore
\begin{equation}
	S(\omega) = 1 - \sqrt{\mathbf{K}_e^T} \left[ G(\omega)^{-1}  \right]_{[kN:(k+1)N, kN:(k+1)N]} \sqrt{\mathbf{K}_e} .
\end{equation}

\subsubsection{Simulation of Bloch oscillation based on the Hamiltonian model}
In the lab frame, we have
\begin{equation}
	\oph = \sum_{n=-\infty}^\infty \omega_n \opad_n \opa_n + 2 \cos (\Omega t) \sum_n (g_n \opad_n \opa_{n+1} + h.c.) .
\end{equation}
In the rotating frame of $\opa_n = \opb_n e^{-i(\omega_0+n\Omega) t}$, we have
\begin{equation}
	\oph = \sum_{n=-\infty}^\infty \Delta_n \opbd_n \opb_n + \sum_n (g_n \opbd_n \opb_{n+1} + h.c.) ,
\end{equation}
where $\Delta_n = \omega_n - \omega_0 - n\Omega$.

At time $t=0$, prepare initial state $\ket{\Psi(t=0)} = \cdots \ket{\beta_{n-1}} \otimes \ket{\beta_n} \otimes \ket{\beta_{n+1}} \cdots$ and denote $\bm{\beta}(0)$ as the vector of coherent-state amplitudes
\begin{equation}
	\bm{\beta}(0) = \begin{pmatrix} \vdots \\ \beta_{n-1} \\ \beta_n \\ \beta_{n+1} \\ \vdots \end{pmatrix} .
\end{equation}
At time $t > 0$, the coherent-state amplitudes are given by
\begin{equation}
	\bm{\beta} (t) = e^{-i(H-i\frac{K}{2})t} \bm{\beta} (0) ,
\end{equation}
where $K=\text{diag}(\cdots, \kappa_n = \kappa_{e,n}+\kappa_{i,n}, \cdots)^T$ includes the total loss rates of all modes. Using the input-output relation, we get the time-domain output field in the lab frame
\begin{equation}
	S_{out} (t) = \sum_n \sqrt{\kappa_{e,n}} \beta_n(t) e^{-i(\omega_0+n\Omega) t} .
\end{equation}
The spectrum of the output field is
\begin{equation}
	S_{out} (\omega) = i \sum_n \sqrt{\kappa_{e,n}} \beta_n(\omega - \omega_0 - n\Omega) ,
\end{equation}
where
\begin{equation}
	\bm{\beta} (\omega) = \left( H - i\frac{K}{2} - \omega I \right)^{-1} \bm{\beta} (0) .
\end{equation}

\subsubsection{Effective onsite potential in the synthetic lattice}

The nonlinear spacing between modes influences the system dynamics. Consider a frequency-modulated magnon system described by
\begin{equation}
H = \sum_n g_n a_n^\dagger a_{n+1} e^{i(\omega_{n+1}-\omega_n-\Omega)t} + a_{n+1}^\dagger a_n e^{-i(\omega_{n+1}-\omega_n-\Omega)t}.
\end{equation}

Applying the gauge transformation
\begin{equation}
a_n(t) = e^{-i(\omega_n - n\Omega)t} \tilde{a}_n(t),
\end{equation}
the Hamiltonian becomes time-independent,
\begin{equation}
\tilde{H} = \sum_n g_n \tilde{a}_n^\dagger \tilde{a}_{n+1} + \text{h.c.} + \sum_n V_n \tilde{a}_n^\dagger \tilde{a}_n,
\end{equation}
with the effective onsite potential
\begin{equation}
V_n = n\Omega - \omega_n.
\end{equation}

The potential gradient between neighboring sites,
\begin{equation}
\Delta_n = V_n - V_{n-1} = \Omega - (\omega_n - \omega_{n-1}),
\end{equation}
acts as an effective static electric field in the synthetic frequency lattice. When the gradient $\Delta_n$ is approximately uniform, the system supports Bloch oscillations along the synthetic frequency dimension. When focusing on a given mode $n$ and its nearest neighbors (which is the case in our experimental system), Bloch oscillations can be approximately achieved by enforcing $|\Delta_n| = |\Delta_{n-1}|$, which yields the modulation frequency
\begin{equation}
\Omega_n = \frac{\omega_{n+1} - \omega_{n-1}}{2}.
\end{equation}

In contrast, for the full synthetic lattice with a linearly decreasing mode spacing $\omega_{n+1} - \omega_n = \Delta_0 - \beta n$, the effective potential $V_n = \omega_n - n\Omega$ gives rise to a position-dependent onsite gradient
\begin{equation}
V_{n+1} - V_n = \Delta_0 - \Omega - \beta n,
\end{equation}
resulting in a nonuniform force and consequently more complex dynamical evolution, which is beyond the scope of this work and will be studied in the future.

\vskip 15pt

\subsection{Micromagnetic model}

\subsubsection{Magnon dispersion at equilibrium}
We consider a uniformly magnetized rectangular film with in–plane dimensions $l=3~\mathrm{mm}$ and $w=2~\mathrm{mm}$ and thickness $d=23~\mu\mathrm{m}$. The saturation magnetization is estimated as $m_s \approx 0.185\ \text{T}$. In the presence of a static bias field $\mathbf H_0=H_0\,\hat{\mathbf z}$ and a longitudinal parametric drive of amplitude $H_1$ and angular frequency $\Omega$ (collinear with $\mathbf H_0$), the magnetization dynamics are governed by the Landau–Lifshitz–Gilbert (LLG) equation
\begin{equation}
\dot{\mathbf m}
= -\gamma\,\mathbf m \times \!\Big[\mathbf H_0 + \mathbf h_d[\mathbf m] + H_1\cos(\Omega t)\,\hat{\mathbf z}\Big]
+ \alpha\,\mathbf m \times \dot{\mathbf m},
\label{eq_s1}
\end{equation}
where $\gamma = 28\ \text{GHz/T}$ is the gyromagnetic ratio and $\alpha\ll1$ is the (dimensionless) Gilbert damping parameter. The magnetostatic (demagnetizing) field generated by dipolar interactions reads as
\begin{equation}
\mathbf h_d(\mathbf r,t)=\int d^3 r'\,\boldsymbol\Gamma(\mathbf r-\mathbf r')\cdot\mathbf m(\mathbf r',t), \label{eq_s1.2}
\end{equation}
where $\boldsymbol\Gamma$ is the magnetostatic Green tensor \cite{guslienko2011magnetostatic}. The relevant modes lie in the magnetostatic, long-wavelength regime, where dipolar contributions dominate and exchange is parametrically small in micron-scale YIG films.
In this regime, the lowest-order thickness mode is nearly uniform along $y$, 
so the full Green tensor can be replaced by its thickness average. 
Since $d \ll \{l, w\}$, variations across the thickness are negligible compared 
to the in-plane dimensions, leading to
\begin{equation}
\Gamma_{\alpha\beta}(\boldsymbol{\rho},\boldsymbol{\rho}')
= \frac{1}{d} \int_0^d \! dy \int_0^d \! dy'\,
\Gamma_{\alpha\beta}(\mathbf{r}-\mathbf{r}'),\label{kernal}
\end{equation}
where $\mathbf r=(\boldsymbol\rho,y)$, with  $\boldsymbol\rho=(z,x)$. Linearizing the LLG equation (\ref{eq_s1}) around the equilibrium configuration, substituting the averaged kernel from Eq.~(\ref{kernal}), and solving the resulting eigenvalue equation in the Fourier domain yield the standard magnetostatic dispersion relation for the fundamental film mode~\cite{jorzick1999brillouin}:
\begin{equation}
\omega_{\mathbf k}=\gamma\sqrt{\Big(H_0+m_s\,\frac{k_x^2}{k^2}f(kd)\Big)
\Big(H_0+m_s\,[1-f(kd)]\Big)} ,
\quad f(kd)=1-\frac{1-e^{-kd}}{kd},
\end{equation}
with $k=\sqrt{k_x^2+k_z^2}$. In our device the loop antenna is aligned parallel to the long edge and excites/detects waves that propagate in–plane perpendicular to $\mathbf H_0$, i.e., Damon–Eshbach–type MSSWs along $\hat{\mathbf x}$ \cite{eshbach1960surface}. That is, we set $k_z=0$ and $k \equiv k_x$. The finite width imposes mixed (Robin) boundary conditions on the transverse
magnetization $m=m_x+im_y$
at $x=\pm l/2$~\cite{jackson2021classical}:
\begin{equation}
\partial_x m|_{x=\pm l/2}=\mp {\zeta_\pm}\,m|_{x=\pm l/2},
\end{equation}
with $\zeta_\pm$ set by the electromagnetic environment at each edge. Within the slab interior \((|y|<d/2)\), modes propagating perpendicular to \(\mathbf{H}_0\) are described by the spatial profile
\begin{equation}
\Phi(y,x)\propto
\begin{cases}
e^{k(d/2-y)} \,e^{i k x}, & k_x=k,\\[3pt]
e^{k(d/2+y)}\,e^{-i k x}, & k_x=-k.
\end{cases}  \label{eq_s8}
\end{equation}
so that opposite propagation directions localize on opposite film surfaces~\cite{eshbach1960surface}. In the magnetostatic interior, Maxwell’s equations reduce to the 2D Laplace equation \(\partial_x^2\Phi+\partial_y^2\Phi=0\), which implies \(k_y^2=-k_x^2\) and thus exponential decay across \(y\)~\cite{hurben1995theory}. However, in the long-wavelength limit \(kd\ll1\), the thickness dependence is weak: the profile is nearly uniform across the thickness, and standing waves can develop along \(x\) \cite{hurben1995theory}. 
 The generalized Robin boundary conditions, together with the mode-profile ansatz 
$\Phi(u) = A\cos(ku) + B\sin(ku)$ (with $u = x + l/2$), yield the following expressions 
for the corresponding mode profiles:
\begin{equation}
\Phi_n(u)=N_n\big(\cos(k_nu)+\frac{\zeta}{k_n}\sin(k_nu)\big), \label{eq_s10}
\end{equation}
where $k_n$  is obtained from the transcendental equation
\begin{equation}
    (\zeta_{+}\zeta_{-}-k^{2})\,\sin(kl)+k(\zeta_{+}+\zeta_{-})\,\cos(kl)=0,
\end{equation}
and $N_n$ is a normalization constant. For $\zeta_+ = -800~\mathrm{m^{-1}}$ and $\zeta_- = 1500~\mathrm{m^{-1}}$, 
the wave numbers are approximately given by $k_n \approx n\pi / l$. The frequency of the lowest mode $\omega_n\equiv\omega_{k_n}$ $\sim 4.067\,\mathrm{GHz}$ at \(H_0 \approx 76.5~\mathrm{mT}\) corresponds to \(n=3\) with \(k_3 \approx 3.14\times10^{3}~\mathrm{m^{-1}}\), yielding \(kd \approx 0.072 \ll 1\), which corroborates the validity of the magnetostatic, thickness-averaged approximation used in this derivation.

\subsubsection{Driven dynamics}

In the presence of the external parametric drive, the linearized LL equation becomes
\begin{equation}
\begin{bmatrix}
\dot{m^x} \\
\dot{m^y}
\end{bmatrix}
= -\gamma 
\begin{bmatrix}
0 & H_0 + H_1 \cos(\Omega t) - m_s \Gamma_{yy}(\bold{k}) \\
- H_0 - H_1 \cos(\Omega t)+ \Gamma_{xx}(\bold{k}) m_s  & 0
\end{bmatrix}
\begin{bmatrix}
m^x \\
m^y
\end{bmatrix}.\label{eq_s11}
\end{equation}

Since the strength of the ac drive \(H_1\) is significantly weaker than the static bias field, we treat \(H_1\cos(\Omega t)\) as a weak perturbation. To first order, the mode frequency is modulated as
\begin{equation}
\omega_n(t)=\omega_n+\left.\frac{\partial\omega_n}{\partial H_0}\right|_{H_0} H_1\cos(\Omega t)
=\omega_n+\Delta\cos(\Omega t),
\label{eq_s12}
\end{equation}
with \(\Delta\equiv \left.\frac{\partial\omega_n}{\partial H_0}\right|_{H_0} H_1\) being the Floquet-induced magnon mode frequency deviation from equilibrium. The parametric drive imposes a periodic, first-order modulation of the normal-mode frequencies, which is taken to be mode independent (i.e., independent of $n$),  and weakly distorts their spatial profiles.
 The profile distortion feeds back through the demagnetizing Green tensor: the modulated magnetization generates oscillatory dipolar fields [Eq.~(\ref{eq_s1.2})], which, in turn, drive the dynamics of the complex transverse amplitude via
\begin{equation}
\dot m(\mathbf r,t)=-i\,\gamma m_s\big[h_d^x(\mathbf r,t)+i\,h_d^y(\mathbf r,t)\big].
\label{eq_s14}
\end{equation}

We expand the transverse magnetization in the eigenmode basis of the undriven system
\begin{equation}
{m}(t) = \sum_n a_n(t) \Phi_n(u),\label{eq_s15}
\end{equation} 
\(a_n(t)\) are slowly varying modal envelopes defined in a frame rotating at \(\omega_n\).  Plugging Eq. (\ref{eq_s15}) in Eq. (\ref{eq_s14}) and invoking the rotating wave approximation leads to
\begin{align}
\dot{a}_m(t) &= - \frac{i \gamma m_s}{2} \sum_{n \neq m} \left( G^x_{nm} + i G^y_{nm} \right)a_n(t). \label{eq_s16}
\end{align}
The coupling coefficients in Eq. (\ref{eq_s16}) are matrix elements of the  thickness- and width–averaged magnetostatic Green tensor~\cite{guslienko2011magnetostatic} in the modal basis,
\begin{equation}
G_{n m}^y = \iint \Phi_m(x)\,\Gamma_{yy}(x-x')\,\Phi_{n}(x')\,dx\,dx',
\qquad
G_{n m}^x = \iint \Phi_m(x)\,\Gamma_{xx}(x-x')\,\Phi_{n}(x')\,dx\,dx',
\label{eq_s17}
\end{equation}
with kernels
\begin{equation}
\begin{aligned}
\Gamma_{yy}(x-x') &= \frac{4}{dl}\Bigg[
l\sinh^{-1}\!\left(\frac{l}{\sqrt{d^2+\xi^2}}\right)
- l\sinh^{-1}\!\left(\frac{l}{\xi}\right)
+ \sqrt{d^2+\xi^2}-\xi \\[3pt]
&\quad
+ \sqrt{l^2+\xi^2}-\sqrt{d^2+l^2+\xi^2}
\Bigg], \\[6pt]
\Gamma_{zz}(x-x') &= \frac{4}{dl}\Bigg[
d\sinh^{-1}\!\left(\frac{d}{\sqrt{l^2+\xi^2}}\right)
- d\sinh^{-1}\!\left(\frac{d}{\xi}\right)
+ \sqrt{l^2+\xi^2}-\xi \\[3pt]
&\quad
+ \sqrt{d^2+\xi^2}-\sqrt{d^2+l^2+\xi^2}
\Bigg], \\[6pt]
\Gamma_{xx}(x-x') &= - \delta(x-x') - \Gamma_{zz}(x-x') - \Gamma_{yy}(x-x'),
\end{aligned}
\label{eq_s18}
\end{equation}
where $\xi=|x-x'|$. The diagonal coupling terms \(n=m\) renormalize the mode frequencies by a parametrically small self-energy shift, which is neglected henceforth. 

In the absence of periodic driving, the laboratory-frame modal amplitudes carry the carrier phase \(e^{-i\omega_n t}\).  
When the frequency is parametrically modulated according to Eq.~(\ref{eq_s12}), the source term on the right-hand side of Eq.~(\ref{eq_s16}) is modified as $a_n(t) \to a_n(t)\, e^{-i  \tfrac{\Delta}{\Omega} \sin \Omega t }$, where the exponential’s frequency argument follows from integrating Eq.~(\ref{eq_s12}) over the interval $0$ to $t$. Expanding $a_m$ in terms of the Fourier harmonics
\begin{align}
a_m(t) &= \sum_k a_m^k e^{ik\Omega t},\label{eq_s19}
\end{align}
and plugging Eq.~(\ref{eq_s19}) into Eq. (\ref{eq_s16}), we obtain
\begin{align}
\sum_k \left( i k \Omega a_m^k + \dot{a}_m^k \right) e^{i k \Omega t} 
&= -i \gamma m_s \frac{1}{2} \sum_{n \neq m} \sum_{\ell} \sum_{r} 
\tilde{G}_{nm} a_n^\ell \mathcal{J}_r\!\left( \frac{\Delta}{\Omega} \right) e^{i (\ell - r) \Omega t}, \label{eq_s20}
\end{align}
where $\tilde{G}_{nm}=G^x_{nm} + i G^y_{nm}$ and we used the Jacobi-Anger expansion
\begin{equation}
e^{-i \left(  \frac{\Delta}{\Omega} \sin(\Omega t) \right)} = \sum_{\ell=-\infty}^{\infty} \mathcal{J}_{\ell}\left( \frac{\Delta}{\Omega}\right) e^{-i (\ell \Omega)t} .
\label{eq_s21}
\end{equation}
where $\mathcal{J}_{\ell}$ denote the Bessel functions of order $\ell$.

Projecting Eq. (\ref{eq_s20}) into $e^{i p \Omega t}$ and equating its coefficients on either side, yields the sideband-resolved dynamical equation
\begin{equation}
\left( i p \Omega \, a_m^p + \dot{a}_m^p \right) 
= -i \gamma m_s \frac{1}{2} \sum_{n \neq m} \sum_{\ell \neq p} 
\tilde{G}_{nm} \mathcal{J}_{\ell - p}\!\left( \frac{\Delta}{\Omega} \right)a_n^\ell \label{eq_s22}.
\end{equation}
Guided by the experimental observations, we restrict the model to nearest-neighbor (NN) couplings in $n$ and first order sidebands.
 Equation (\ref{eq_s22}) then becomes
\begin{equation}
    \left( i p \Omega \, a_m^p + \dot{a}_m^p \right)=-ig_{m,m+1}\!\left[a_{m+1}^{p-1}+a_{m+1}^{p+1}\right]
-i g_{m-1,m}\!\left[a_{m-1}^{p-1}+a_{m-1}^{p+1}\right],\label{eq_s22.1}
\end{equation}
where $g_{m,m+1}\approx \frac{\gamma m_s}{2}\tilde{G}_{m,m+1}\mathcal{J}_1(\Delta/\Omega)$. 

This restriction to nearest-neighbor coupling is justified by the strong suppression of higher-order processes in our experimental regime. The ratio between next-nearest-neighbor (NNN) and nearest-neighbor coupling strengths scales as $r = J_2(\Delta/\Omega)/J_1(\Delta/\Omega)$. For our experimental parameters with $\Delta/\Omega \approx 0.4$, this ratio is approximately $r \approx 0.1$, meaning NNN coupling is an order of magnitude weaker than NN coupling. The good agreement between the nearest-neighbor tight-binding model [Eq.~(2) in main text] and experimental observations in Fig.~3 provides further validation of this approximation. 
 
Summing Eq.~(\ref{eq_s22.1}) over \(p\)  and restoring the carrier phase yields
\begin{equation}
i\,\dot a_m(t)=\omega_m a_m(t)
+ 2\cos(\Omega t)\,\Big[\, g_{m,m+1}\,a_{m+1}(t)+g_{m-1,m}\,a_{m-1}(t)\,\Big].
\end{equation}
This corresponds to dynamics governed by the following time-dependent tight-binding Hamiltonian, in which we redefine $g_n \equiv g_{n,n+1}$:
\begin{equation}
\hat H=\sum_{n=1}^{N}\omega_n\,\hat a_n^\dagger\hat a_n
+\sum_{n=1}^{N-1}\Big[\,2\,g_{n,n+1}\cos(\Omega t)\,\hat a_n^\dagger\hat a_{n+1}+\text{h.c.}\,\Big].
\end{equation}

\section{Device and experimental setup}

\subsection{Detailed device parameters}

\subsubsection{Device description}
The device used in our experiment is a piece of single-crystalline yttrium iron garnet (YIG) film. The YIG film is epitaxially grown (single sided) on a 500\,$\mu$m-thick gadolinium gallium garnet (GGG) substrate. The YIG thickness is 23\,$\mu$m, with a lateral dimension of 2 mm $\times$ 3 mm. The YIG chip is biased along its short edge by a piece of permanent magnet, whose position can be moved to tune the strength of the bias magnetic field. A coaxial loop antenna is positioned near the surface of the YIG chip to selectively excite and detect magnons, with the loop aligned parallel to the long edge of the YIG piece. Under this configuration, the propagation direction of the magnon modes is in plane and parallel to the long edge, which is perpendicular to the bias magnetic field direction. As a result, the magnon modes are magnetostatic surface waves (MSSWs). Due to the finite chip size, these MSSW magnons form a series of discrete standing wave modes.

In the device reflection spectrum measured using a vector network analyzer (VNA), the extinction ratio of the magnon modes decreases with increasing mode number, which is attributed to reduced coupling efficiency between the loop antenna and higher-order magnon modes. These spectral features confirm that the observed modes are MSSWs rather than backward volume magnetostatic waves (MSBVWs), which exhibit negative dispersion slopes and typically show higher-order modes (at higher frequencies) with larger extinction ratios. For MSSWs, the mode order increases with frequency, and the mode distribution becomes denser. But for MSBVWs, the mode order increases when the frequency decreases, and the mode distribution is denser at lower frequencies. From our experimental observation, the spectral signatures of the measured magnon modes match that of the MSSWs.

For the five-mode synthetic lattice demonstrated in the main text, the strong-coupling condition ($g_n > \kappa_{\text{avg}}$) is satisfied up to Mode 4. Specifically, $g_3/2\pi = 2.07$ MHz exceeds the average dissipation rate of Modes 3 and 4. The coupling between Modes 4 and 5 is measured to be $g_4/2\pi = 1.15$ MHz, while the corresponding average dissipation rate is $\kappa_{\text{avg}}/2\pi = 1.32$ MHz, placing this coupling just outside but very close to the strong-coupling regime. For higher-order modes, the coupling strength progressively decreases while dissipation rates remain in the range of 1–1.5 MHz.

The interaction strength can be enhanced by increasing the saturation magnetization $m_s$, improving mode overlap $G_{n,n+1}$  through optimized device geometry or boundary conditions, or increasing the modulation field strength $H_1$ (as indicated by the $J_1(\Delta/\Omega)$ factor in Eq.~(3) of the main text). For example, doubling $H_1$ leads to an approximate doubling of the coupling strength, provided $\Delta/\Omega < 1$. Additionally, Ref.~[55] of the main text demonstrates that the LC-resonance enhancement remains robust even when the LC resonance is detuned from the mode spacing.

In our experiments, only first-order sidebands ($l = \pm 1$) of each mode are considered for simplicity and clarity. However, second-order sidebands are also visible in the spectra, which show up Fig.\,3 of the main text as the unlabeled sidebands with much smaller amplitudes. This suggests that longer-range coupling is possible in our synthetic dimension system, but it lies beyond the scope of this work and will be explored in future studies.

\subsubsection{Floquet drive power and thermal effect}
In our experiment, the Floquet drives are amplified using a high-gain amplifier with a gain of 40\,dB. A Floquet drive with an input amplitude of $V_F = 200$\,mV corresponds to an output power of approximately 32.5\,dBm after amplification, just before being delivered to the coil. Based on the measured extinction ratio of the LC resonance at around 30 MHz, which is around 10 dB, the estimated power delivered to the LC resonance is 32 dBm. But note that this value corresponds to the ideal case. In real devices, the LC resonator will be heated up by the driving power, causing spectral shift of the resonance. As a result, the on-resonance drive signal will become off-resonance and its power will decrease from the ideal case.

At high drive amplitudes, the heat generated by the coil significantly shifts both the magnon and LC resonances, thereby obscuring the synthetic dimension effects. To mitigate this, we limit the drive amplitude to below 200~mV in our measurements. Even at these reduced drive levels, thermal effects persist, leading to resonance shifts and a reduction in the effective Floquet drive power. To minimize the impact of thermal drift on the magnon resonances, a wait time of 5 to 30 seconds is introduced between successive points in the parameter sweep.

\subsection{Measurement setup and procedure}
\subsubsection{VNA spectrum and mode profile measurement} 
The magnon resonance is measured using a VNA via reflection (S11) measurement. To suppress interferences, an circulator is used to separate the input and output signal to the YIG device. Therefore, the Src. 1 and detector in Fig.\, 1(a) of the main text are the two ports of the VNA. In addition, the LC resonance is also first characterized using the VNA before Floquet drive signals are sent to the device, which allows us to adjust the LC resonance frequency to desired values by tuning the variable capacitor.

The spatial mode profiles shown in the inset of Fig.~1(c) are measured by scanning a coaxial loop probe over the surface of the YIG film using automated $xyz$ stages, while monitoring the reflection coefficient $S_{11}$ using a VNA. The probe is similar to the one used for magnon excitation and detection in the main experiment. At each position, a full $S_{11}$ spectrum is recorded as a function of frequency, where the magnon modes appear as resonance dips similar to those shown in Fig.~1(c). To obtain the spatial profile of a given mode, the $S_{11}$ magnitude at the corresponding resonance frequency is extracted and plotted as a function of probe position across the entire YIG film surface. This measurement characterizes the intrinsic magnon mode profiles of the YIG film and is performed without the Floquet drive coil.

\subsubsection{Time-resolved band structure spectroscopy} 
\begin{figure}[tb]
\includegraphics[width=0.5\linewidth]{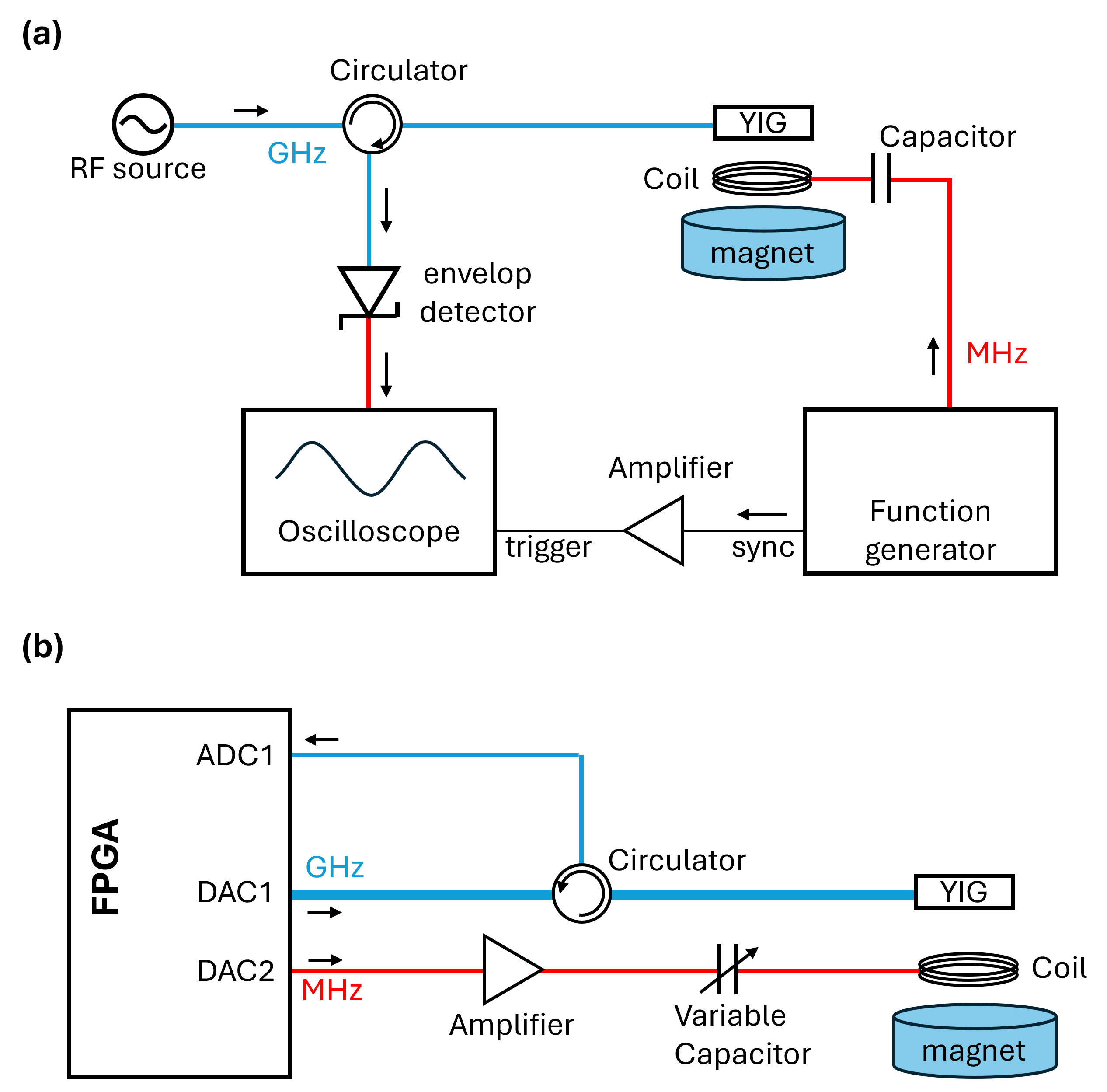}
\caption{(a) Schematics of the setup for Time-resolved band structure spectroscopy. (b) Schematics of the Bloch oscillation measurement setup.}
\label{figSM1}
\end{figure}

To measure the Floquet band structure, a single-tone microwave signal is applied to the device, and the reflected output is measured using an envelope detector and an oscilloscope, as schematically shown in Fig.\,\ref{figSM1}(a). By sweeping the microwave frequency, the reflected power can be recorded as a function of frequency, yielding a spectrum analogous to that obtained from a VNA, especially in the absence of the Floquet drive, where the spectrum remains stationary and the magnon mode shows up in the spectrum as a dip at a fixed frequency. When the Floquet drive is activated, the magnon resonance will shift over time, and this measurement method enables temporal probing of the magnon mode frequency. By synchronizing the microwave signal—converted to a voltage via the envelope detector—with the Floquet drive signal on the oscilloscope, the real-time spectrum at different phases of the Floquet oscillation can be observed. 

Specifically, for each time-trace measurement, the microwave tone is fixed at a given frequency, and the reflected power is recorded as a function of time using the oscilloscope. This process is repeated for different microwave frequencies. The resulting two-dimensional sweep over frequency and time yields the measured spectra shown in Fig.\,2 of the main text.

\subsubsection{Bloch oscillation measurement}

The Bloch oscillation measurement setup is shown in Fig.~S1(b). The pulse sequences are controlled by an FPGA board. The GHz excitation pulse at frequency $\omega_e = \omega_4$ is generated by DAC~1, sent through a circulator to the YIG device, and the reflected signal is routed back through the circulator to ADC~1 for IQ demodulation. The MHz Floquet drive pulses are generated by DAC~2, amplified, and delivered to the coil via the LC resonator circuit.

Mode 4 is selectively excited by an RF pulse at frequency $\omega_e = \omega_4$, following the pulse sequence shown in Fig.~4(a). The Floquet drive at frequency $\Omega/2\pi = 30$~MHz is then applied, during which the reflected signal at $\omega_4$ is continuously recorded by the FPGA via IQ demodulation at the same frequency $\omega_4$ throughout the entire Floquet drive window. The demodulated signal amplitude is recorded for four different drive amplitudes: $V_F = 0$, 110, 130, and 150~mV (before amplification), yielding the time traces plotted in Fig.~4(c). The observation window is limited to 0.5~$\mu$s by the finite magnon lifetime and the detection noise floor.

\begin{figure}[tb]
\includegraphics[width=0.6\linewidth]{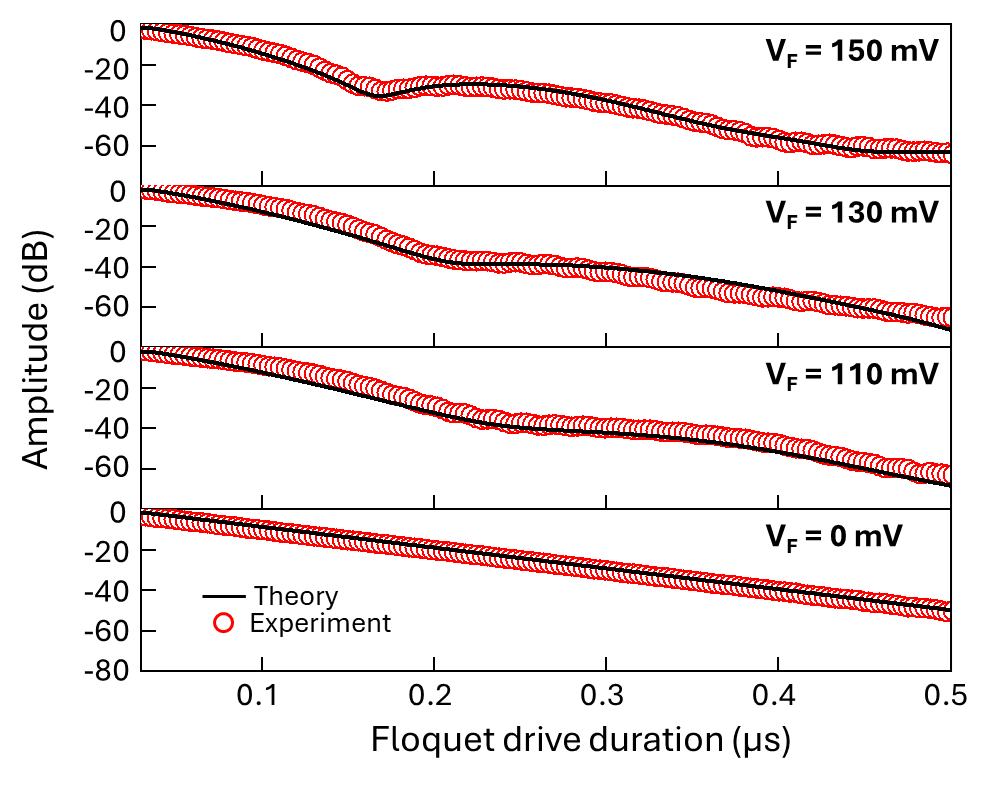}
\caption{Complete time-resolved reflection data for the Bloch oscillation measurement, showing the demodulated signal amplitude for $V_F =0$, 110, 130, 150~mV. The $V_F = 0$ curve serves as the exponential decay background that is removed to reveal the oscillatory dynamics shown in Fig.~4(c) of the main text.
}
\label{figSM2}
\end{figure}

The $V_F = 0$ curve, which exhibits pure exponential decay due to magnon dissipation in Mode 4, serves as the background reference, as shown in Fig.~\ref{figSM2}. This background is divided out from the three driven curves ($V_F = 110$, 130, 150~mV) to reveal the oscillatory dynamics shown in the right panel of Fig.~4(c) of the main text.

After background removal, a residual decaying envelope remains in the normalized signal. This residual arises because the Floquet drive redistributes population across Modes 3, 4, and 5, each with slightly different dissipation rates ($\kappa_3/2\pi = 1.70$~MHz, $\kappa_4/2\pi = 1.29$~MHz, $\kappa_5/2\pi = 1.35$~MHz). The effective decay rate of the Mode 4 signal under modulation therefore deviates from the single-mode $\kappa_4$ used for background removal, leaving a residual envelope. Notably, this residual decay is itself a direct signature of Floquet-induced inter-mode coupling---in the absence of coupling, the background removal would be perfect and no residual envelope would remain. Simulations confirm that this residual decay is not caused by dephasing from nonuniform mode spacing, but rather by population redistribution across modes with different dissipation rates.

\end{document}